\documentclass[twocolumn,5p,nonatbib,final,times,9pt]{elsarticle}
\usepackage{graphics} % for pdf, bitmapped graphics files
\usepackage{framed,enumitem} 
\usepackage{epsfig} % for postscript graphics files
\usepackage{arydshln}
\usepackage{times} % assumes new font selection scheme installed
\usepackage{amsmath} % assumes amsmath package installed
\usepackage{amssymb}  % assumes amsmath package installed
\usepackage[noadjust]{cite}
\usepackage{xcolor}
\newcommand{\FV}[1]{{\color{black}#1}}

\journal{Journal of mechatronics}
\begin{document} 
\begin{frontmatter}
\title{
	Active Compensation of Position Dependent Flexible Dynamics\\ in High-Precision Mechatronics
}%

\author[TUe]{Yorick Broens\corref{cor1}}
\ead{Y.L.C.Broens@tue.nl}
\author[TUe,ASML]{Hans Butler}
\author[ASML]{Ramidin Kamidi}
\author[ASML]{Koen Verkerk}
\author[TUe]{Siep Weiland}

\address[TUe]{Control Systems Group, Eindhoven University of Technology, Eindhoven, The Netherlands}
\address[ASML]{ASML, Veldhoven, The Netherlands}

\cortext[cor1]{Corresponding author}
\begin{abstract} 
Growing demands in the semiconductor industry {necessitate} increasingly stringent requirements on throughput and positioning accuracy of lithographic equipment. {Meeting these demands involves employing highly aggressive motion profiles, which introduce position-dependent flexible dynamics, thus compromising achievable position tracking performance. This paper introduces a control approach enabling active compensation of position-dependent flexible dynamics by extending the conventional \FV{rigid-body} control structure to include active control of flexible dynamics. To facilitate real-time implementation of the control algorithm, appropriate position-dependent weighting functions are introduced, ensuring computationally efficient execution of the proposed approach. The efficacy of the proposed control design approach is demonstrated through experiments conducted on a state-of-the-art extreme ultraviolet (EUV) wafer stage.}
\end{abstract}

\begin{keyword}
	Estimation \sep Motion Control Systems \sep Observers \sep Parameter-Varying Systems
	%% keywords here, in the form: keyword \sep keyword
	
	%% PACS codes here, in the form: \PACS code \sep code
	
	%% MSC codes here, in the form: \MSC code \sep code
	%% or \MSC[2008] code \sep code (2000 is the default)
	
\end{keyword}
\end{frontmatter}

\begin{section}{Introduction}
\label{Section:Introduction}
Using a high-precision lithographic process, {the} production of integrated circuits is realized by projecting extreme ultraviolet light through a blueprint of the pattern {onto} a silicon wafer using projection optics. {These optics shrink and focus the pattern onto the photosensitive silicon wafer}. {To} achieve high throughput and high reliability, the wafer is positioned under the projection optics using the wafer stage module, { a planar motor system capable of achieving nanometer accuracy in position tracking. } {Growing} demands in the semiconductor industry {impose} increasingly stringent requirements on throughput and position tracking performance, {aligning with the challenging expectations of Moore's law, see \cite{ASMLpaper}.} {Highly aggressive motion profiles with accelerations up to 5g are required to meet these demands, leading to high-frequency position tracking errors due to the limited stiffness of the mechanical structure, see \cite{6518502,190Steinbuch}.
Additional complications are introduced by relative position measurements of the moving-body, necessitating for rigid-body coordinate transformations to accurately relate the point of control on the moving-body to the position measurements, see \cite{Murray-book}. As effective as this approach is in decoupling (position dependent) rigid-body dynamics, high frequency position dependent coupling persists in the shape of position dependent resonance dynamics, forcing the corresponding \emph{linear-time-invariant} (LTI) controller to enforce robustness at the cost of closed-loop performance.
%However, this approach introduces position-dependent flexible dynamics, forcing the corresponding \emph{linear-time-invariant} (LTI) controller to handle them in terms of robustness at the cost of closed-loop performance.}
Several studies have proposed extensions of the conventional rigid-body control framework towards active control of flexible dynamics by introducing an output-based modal state observer which is capable of reconstructing the flexible dynamics, see  \cite{a059727d0f484157afa82739a695e87e,7158898,inproceedings,43ba17bf4632486ea6a2131be6b8bd5e}. The reconstructed modal states of the flexible modes can then be further utilized for active control of \FV{resonance modes} by introducing a state feedback (PD-type of controller). Nonetheless, design and implementation of such a flexible mode controller is complicated by the position dependent nature of the flexible dynamics. Additional practical complications are introduced by the ultra-high sampling frequency, \FV{i.e.,} 20 kHz, which severely limits practical feasibility of the attainable motion control possibilities.}
%This observer allows for the reconstruction of flexible modes of the moving body, enabling active manipulation of its modal parameters. However, the design and implementation of this approach are complicated by the position-dependent nature of the flexible dynamics and the ultra-high sampling frequency of the system, \FV{i.e.,}, 40kHz, which severely limits feasible control possibilities.}

{To address the aforementioned limitations, a novel position dependent control design and implementation approach are presented. The proposed approach exploits the position-dependent nature of the flexible dynamics by introducing suitable position-dependent weighting functions for computationally efficient reconstruction of the modal state dynamics of the moving-body. Additionally, a state feedback design approach is presented, which allows for active control of resonance modes, ensuring desirable closed-loop performance.}

The main contributions of this paper are:
\begin{itemize}
 \item[(C1)] {Development of novel position dependent output-based modal observer design using position dependent weighting functions, allowing for real-time application of the proposed approach on physical hardware.}
  \item[(C2)] {Extension of the conventional rigid-body control structure towards active position dependent control of flexible dynamics through an output-based modal observer.}
\end{itemize}

This paper is {structured} as follows. First, the problem formulation is presented in Section \ref{Section:ProblemFormulation}. Next, Section \ref{Section:ControlDesignApproach} presents the design of the flexible mode control loop. Section \ref{ExperimentalVAlidationSection} details the experimental validation of the proposed approach on a state-of-the-art EUV wafer stage. Section \ref{Section:Conclusion} draws conclusions on the presented work.

\end{section}
%%%%%%%%%%%%%%%%%%%%%%%%%%%%%%%%%%%%%%%%%%%%%%%%%%%%%%%%%%%%%%%%%%%%%5

\begin{section}{Problem Formulation}
\label{Section:ProblemFormulation}

\begin{subsection}{Background}
{Many high-precision motion systems exhibit position-dependent effects arising from relative position measurements of the moving body, see \cite{176Butler}. To accurately establish a relationship between the point of control on the moving-body and the actual position measurements, position dependent rigid-body coordinate frame transformations are required. To effectively capture these position dependent effects, such systems are often represented in \emph{linear-parameter-varying} (LPV) form, see \cite{5714737}. Consider the equations of motion for a \emph{multiple-input multiple-output} (MIMO) high-precision motion system, which demonstrates position-dependent effects due to relative position measurements of the moving body:
\begin{equation}
P(p(t)):= \begin{cases}
    \begin{split}
M\Ddot{x}(t)&=-D\dot{x}(t)-Kx(t) + \Phi_a u(t) \\
y(t) &= \Phi_s(p(t)) x(t)
\end{split}
\end{cases}
\label{SectionII_Eq_1}
\end{equation}

\noindent 
Here, $M$, $D$, and $K$ $\in \mathbb{R}^{n_x \times n_x}$ represent real symmetric mass, damping, and stiffness matrices respectively. $\Phi_a \in \mathbb{R}^{n_x \times n_u}$ maps the forces acting on the moving body, $u(t)$, to its point of control. $\Phi_s(p(t))\in \mathbb{R}^{n_y \times n_x}$ corresponds to a position-dependent mapping of the point of control on the moving-body to the position measurements, $y(t)$, where the scheduling vector is given by $p(t):\mathbb{R}\rightarrow \mathbb{P} \subseteq\mathbb{R}^{n_p}$, see \cite{176Butler}. To facilitate individual control of the mechanical degrees of freedom (DoF), the motion dynamics given by (\ref{SectionII_Eq_1}) are typically represented in modal form. This involves a state transformation $x(t)=\tilde{V}q(t)$ using the mass-normalized eigenvector matrix $\tilde{V}=M^{-\frac{1}{2}}V$, derived from the characteristic dynamical equation $KV = MV\Lambda$, see  \cite{Gawronski}. The equations of motion in modal form correspond to:
\begin{equation}
P(p(t)):= \begin{cases} \begin{split}
    \Ddot{q}(t)&=-2Z\Omega\dot{q}(t)-\Omega^2q(t) + \tilde{V}^\top\Phi_a u(t) \\
y(t) &= \Phi_s(p(t)) \tilde{V}q(t)
\end{split}
\end{cases},
\label{EqMOtionSectionII}
\end{equation}

\noindent 
where $Z\in \mathbb{R}^{n_x\times n_x}$ is a diagonal matrix containing modal damping parameters, and $\Omega \in \mathbb{R}^{n_x \times n_x}$ is a diagonal matrix containing the system's eigenfrequencies. A secondary state transformation facilitates grouping of the states per mode, \FV{i.e.,} $q(t) = T(q_{_{\mathrm{RB}}}^\top(t) \ q_{_{\mathrm{FM}}}^\top(t))^\top$, where $T$ is given by:
\begin{equation*}
T=\left(I_{n_x \times n_x}\otimes \begin{pmatrix}1 & 0 \end{pmatrix}^\top \ \ I_{n_x \times n_x}\otimes \begin{pmatrix}0 & 1 \end{pmatrix}^\top \right)
\end{equation*}

\noindent Note that $\otimes$ corresponds to the \emph{Kronecker product}. Moreover, the resulting partitioned state-space representation is given by:
\begin{equation}
        P(p(t)) :=
         \left ( \begin{array}{cc|c}
             A_{_{\mathrm{RB}}}  & 0 &B_{_{\mathrm{RB}}}  \\  0& A_{_{\mathrm{FM}}}  & B_{_{\mathrm{FM}}}  \\ 
             \hline 
             C_{_{\mathrm{RB}}}(p(t))   &C_{_{\mathrm{FM}}}(p(t))  & 0
        \end{array}\right),
    \label{SectionII_Eq_2}
\end{equation}

\noindent where $(\cdot)_{_{\mathrm{RB}}}$ are the system matrices that correspond to the rigid body dynamics and $(\cdot)_{_{\mathrm{FM}}}$  correspond to those of the flexible modes. To simplify motion control design procedure, (position dependent) rigid-body decoupling strategies are typically applied, see \cite{198Steinbuch,9993406}. For a system that is of the form (\ref{SectionII_Eq_2}), this can be achieved by introducing the decoupling matrices:
\begin{equation*}
\begin{split}
    T_\mathrm{u}&= \begin{pmatrix}\begin{pmatrix}I_{n_{_{\mathrm{RB}}} \times n_{_{\mathrm{RB}}}} \otimes \begin{pmatrix}0 & 1 \end{pmatrix} \end{pmatrix}B_{_{\mathrm{RB}}}\end{pmatrix}^\dagger, \\
    T_\mathrm{y}(p(t))&= \begin{pmatrix} C_{_{\mathrm{RB}}}(p(t))\begin{pmatrix}I_{n_{_{\mathrm{RB}}} \times n_{_{\mathrm{RB}}}} \otimes \begin{pmatrix}1 & 0 \end{pmatrix} \end{pmatrix}^\top \end{pmatrix}^{\dagger},
    \end{split}
\end{equation*}

\noindent where $n_{_{\mathrm{RB}}}$ corresponds to the number of rigid-body modes of the system. Furthermore, the state-space representation of the rigid-body decoupled system, \FV{i.e.,} $\tilde{P}(p(t))=T_\mathrm{y}(p(t)) P(p(t)) T_\mathrm{u}$, is given by:
\begin{equation}
\begin{small}
       \tilde{P}(p(t)) := 
         \left ( \begin{array}{cc|c}
             A_{_{\mathrm{RB}}}  & 0 &I  \\  0& A_{_{\mathrm{FM}}}  & B_{_{\mathrm{FM}}}T_\mathrm{u}  \\ 
             \hline 
             I   &T_\mathrm{y}(p(t))C_{_{\mathrm{FM}}}(p(t))  & 0
        \end{array}\right)
\end{small}
    \label{SectionII_Eq_3}
\end{equation}

\noindent An important observation that is made is that if the position is constant, \FV{i.e.,} $p(t)= \tt{p}$ for all $t \in \mathbb{R}$, \eqref{SectionII_Eq_3} becomes an LTI system, often referred to as \emph{local dynamics} of the LPV system. Additionally, the form of (\ref{SectionII_Eq_3}), it is observed that introduction of the decoupling matrices removes low-frequency (position dependent) channel interaction of the system as the channels are diagonalized. Nonetheless, high-frequency position dependent interaction of the flexible modes persists, limiting the attainable performance of the system as the rigid-body feedback controller must be designed in a robustified manner. To keep up with industry's challenging expectations, this must be addressed.

To address the issue of flexible dynamics negatively impacting position tracking accuracy while also alleviating the constraints imposed on rigid-body control design, we propose the integration of a flexible mode control loop, consisting of an output-based modal state observer and a state feedback, in combination with the conventional rigid-body control configuration. This combined approach enables active manipulation of the plant, as perceived by the rigid-body feedback controller, thereby allowing for enhanced closed-loop performance while simultaneously mitigating the influence of flexible dynamics.}

{
\subsection{Problem statement}
The problem that is being addressed in this paper is to design and implement a flexible mode controller, such that the contribution of critical position dependent resonance modes to the position tracking error is reduced, while simultaneously relaxing the rigid-body feedback control design constraints, allowing for improved closed-loop performance of lithographic equipment.}

\section{Active control of position dependent resonance dynamics}
\label{Section:ControlDesignApproach}

This section presents an enhancement to the conventional rigid-body control framework aimed at actively managing flexible dynamics. This enhancement involves incorporating an output-based modal observer along with state feedback, enabling active manipulation of the modal plant dynamics. The control interconnection depicted in Figure \ref{fig:Figure1_ASML} is considered, with $\FV{K_{\mathrm{RB}}} \in \mathbb{R}^{n_{{\mathrm{RB}}} \times n_{{\mathrm{RB}}}}$ representing the rigid-body feedback controller, $K_{\mathrm{FF}} \in \mathbb{R}^{n_{{\mathrm{RB}}} \times n_{{\mathrm{RB}}}}$ denoting the rigid-body feedforward controller, and $K_{\mathrm{FM}} O(p(t)) \in \mathbb{R}^{{n_{\mathrm{FM}}}\times n_{{\mathrm{RB}}}}$ governing the active regulation of flexible modes, where ${n_{_{\mathrm{FM}}}}$ signifies the number of flexible modes to be controlled. It's worth noting that this extension of the rigid-body control framework involves breaking down the flexible mode control loop into a position-dependent filter $O(p(t))$ and a static state feedback $K_{\mathrm{FM}}$, thereby \FV{forming a structured position dependent flexible mode controller}.

\begin{figure}[t]
    \centering
    \includegraphics[width=\linewidth]{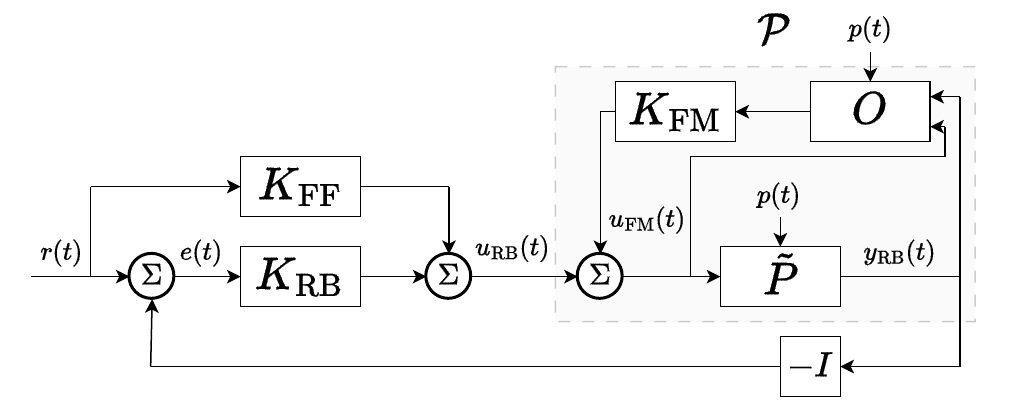}
    \caption{Proposed extension of the rigid-body control interconnection to allow for active regulation of position dependent flexible dynamics.}
    \label{fig:Figure1_ASML}
\end{figure}
%\subsection{Position dependent output-based modal observer design}
For the construction of the output-based modal observer $O(p(t))$, modal truncation strategies are applied to $\tilde{P}(p(t))$, thus minimizing the computational expense of the position-dependent state observer. Furthermore, \eqref{SectionII_Eq_3} is partitioned as:
\begin{equation*}
        \tilde{P}(p(t)) :=
         \left ( \begin{array}{cc:c|c}
             A_{_{\mathrm{RB}}}  & 0 & 0&I  \\  0& A_{_{\mathrm{FM}}}^r & 0  & \tilde{B}_{_{\mathrm{FM}}}^r\\
             \hdashline 
             0 &0 & A_{_{\mathrm{FM}}}^d & \tilde{B}_{_{\mathrm{FM}}}^d
             \\ 
             \hline 
             I   &\tilde{C}_{_{\mathrm{FM}}}^r(p(t)) &\tilde{C}_{_{\mathrm{FM}}}^d(p(t))  & 0
        \end{array}\right),
    \label{SectionIII_Eq_1}
\end{equation*}

\noindent where $(\cdot)_{_{\mathrm{FM}}}^r$ denote the system matrices preserved with respect to the output-based modal observer and $(\cdot)_{_{\mathrm{FM}}}^d$ corresponds to the system matrices discarded with respect to the output-based modal observer. The resulting truncated model for observer design, $\hat{P}(p(t))$, is given by:
\begin{equation}
\begin{small}
       \hat{P}(p(t)) := 
         \left ( \begin{array}{cc|c}
             A_{_{\mathrm{RB}}}  & 0 &I  \\  0& A_{_{\mathrm{FM}}}^r  & \tilde{B}_{_{\mathrm{FM}}}^r \\ 
             \hline 
             I   &\tilde{C}_{_{\mathrm{FM}}}^r(p(t))  & -\tilde{C}_{_{\mathrm{FM}}}^d(p(t)){A_{_{\mathrm{FM}}}^d}^{-1}\tilde{B}_{_{\mathrm{FM}}}^d
        \end{array}\right),
\end{small}
    \label{SectionIII_Eq_1}
\end{equation}

\noindent where the feed-through term corresponds to the compliance correction. It is important to note that the compliance correction is position-dependent due to its dependency on the discarded high-frequency resonance dynamics. Moreover, using \eqref{SectionIII_Eq_1}, the discrete-time observer model is constructed as:
\begin{equation}
             \begin{split}
             \hat{q}(k+1|k) &= \begin{bmatrix}\mathcal{A}-L(p(k))\mathcal{C}(p(k)) \end{bmatrix}\hat{q}(k|k-1)  \\ &+\begin{bmatrix}
            \mathcal{B}-L(p(k))\mathcal{D}(p(k)) & L(p(k)) \end{bmatrix}\begin{bmatrix}
             \tilde{u}(k) \\ y_{\mathrm{RB}}(k)
             \end{bmatrix}
         \end{split},
        \label{SectionIII_Eq_2}
\end{equation}

\noindent where, $\tilde{u}(k)=u_{\mathrm{RB}}(k)+u_{\mathrm{FM}}(k)$, and:
\begin{equation*}
    \begin{split}
        \mathcal{A} &= e^{\mathrm{diag}(A_{_{\mathrm{RB}}}, \ A_{_{\mathrm{FM}}}^r ) T_s},\\
        \mathcal{B} &= 
        \begin{small}
            \mathrm{diag}(A_{_{\mathrm{RB}}}, \ A_{_{\mathrm{FM}}}^r )^{-1}\begin{pmatrix}
             e^{\mathrm{diag}(A_{_{\mathrm{RB}}}, \ A_{_{\mathrm{FM}}}^r ) T_s} -I
        \end{pmatrix}\begin{pmatrix}
            I &{\tilde{B}{_{\mathrm{FM}}}^r}^\top\end{pmatrix}^\top \end{small}, \\ 
            \mathcal{C}(p(k)) &= \begin{small} (
                I \ \ \ \ \  \tilde{C}_{_{\mathrm{FM}}}^r(p(k))  
           )\end{small},\\
            \mathcal{D}(p(k)) &=  \begin{small}-\tilde{C}_{_{\mathrm{FM}}}^d(p(k)){A_{_{\mathrm{FM}}}^d}^{-1}\tilde{B}_{_{\mathrm{FM}}}^d\end{small},
    \end{split}
\end{equation*}

\noindent $T_s$ is the sampling time and $L(p(k))$ corresponds to the \emph{Luenberger} observer gain.
Equation \eqref{SectionIII_Eq_2} prompts several key observations. Firstly, it reveals that the observer dynamics are presented in a one-step-ahead predictor form. This structure is essential to counteract the inherent one-time sample delay introduced by the discrete-time implementation of the modal observer within the rigid-body control framework. Secondly, it is evident that the stability of the observer hinges upon the position-dependent observer gain. To devise this gain, we employ a collection of local truncated plant dynamics $\lbrace\hat{P}_i \rbrace_{i=1}^n$ to generate a corresponding set of local observer gains $\lbrace L_i \rbrace_{i=1}^n$ using the infinite-horizon discrete algebraic Riccati equation, as detailed in \cite{pappas1980numerical}. Here, $n$ represents the number of \emph{local positions} considered for constructing the set of local \FV{LTI} observers $\lbrace O_i \rbrace_{i=1}^{n}$. To construct a position-dependent observer $O(p(k))$, as expressed in \eqref{SectionIII_Eq_2}, we introduce a novel implementation approach. This method computes the weighted sum of the outputs from local observers in a position-dependent manner. Specifically, we approximate the one-step ahead prediction of the to-be-controlled flexible modal state of interest through a position dependent weighted sum of the local observer outputs:
\begin{equation}
    \hat{q}(k+1|k) \approx \sum_{i=1}^n W_i(p(k))  \hat{q}_i(k+1|k),
\label{SectionIII_eq4}
\end{equation}

\noindent where the set of weighting functions $\lbrace W_i(p(k)) \rbrace_{i=1}^n$ are designed to have a polynomial dependency on the scheduling vector $p(k)$. This choice is made feasible by leveraging the spatial characteristics of dynamical resonance modes. Furthermore, the local set of weighting filters takes the following form:
\begin{equation}
    \lbrace W_i(p(k)) \rbrace_{i=1}^n = \left\lbrace \sum_{v=0}^{m_x} \sum_{w=0}^{m_y} q_x^v(k) q_y^w(k) \theta_{i}^{vw}  \right\rbrace_{i=1}^n,
    \label{SectionIII_eq6}
\end{equation}

\noindent where $q_x(k), q_y(k) \in \mathbb{R}^{n_p}$ correspond to the measured $x,y$ position of the moving-body respectively. The parameters $m_x$ and $m_y$ \FV{are tuning parameters which} denote the assumed dependency on the scheduling vector, governing the smoothness of the allowed variation. Moreover, from \eqref{SectionIII_eq6} it is observed that the polynomial weighting function is composed of spatial coordinates, \FV{i.e.,} $q_x(k)$ and $q_y(k)$, and weighting coefficients $\lbrace \theta_{i}^{vw}\rbrace_{i=1}^n$. This structure facilitates a reformulation of the weighting function parameterization as:
\begin{equation}
    \lbrace W_i(p(k)) \rbrace_{i=1}^n = \chi(p(k))\lbrace \Theta_i\rbrace_{i=1}^n,
    \label{SectionIII_eq5}
\end{equation}

\noindent where the spatial vector $\chi(p(k)) \in \mathbb{R}^{1\times (m_x m_y)}$ corresponds to:
\begin{equation*}
    \chi(p(k)) = \begin{pmatrix}
        q_x^0(k) & \hdots & q_x^{m_x}(k)
    \end{pmatrix} \otimes \begin{pmatrix}
        q_y^0(k) & \hdots & q_y^{m_y}(k)
    \end{pmatrix},
\end{equation*}

\noindent and the coefficient vectors $\lbrace \Theta_i \rbrace_{i=1}^n \in \mathbb{R}^{m_x m_y \times 1}$ are given by:
\begin{equation*}
    \lbrace \Theta_i \rbrace_{i=1}^n = \left\lbrace \begin{pmatrix} \theta_i^{00} & \hdots & \theta_i^{m_x \FV{m_y}} \end{pmatrix}^\top \right\rbrace_{i=1}^n
\end{equation*}

\noindent 
Furthermore, substitution of \eqref{SectionIII_eq5} in \eqref{SectionIII_eq4} results in:
\begin{equation}
    \hat{q}(k+1|k) \approx \sum_{i=1}^n\begin{pmatrix}
        \hat{q}_i(k+1|k) \otimes \chi(p(k)) 
    \end{pmatrix}\Theta_i,
    \label{SectionIII_eq9}
\end{equation}

\noindent 
where \eqref{SectionIII_eq9} is linear in the weighting filter coefficients $\lbrace\Theta_i\rbrace_{i=1}^n$, enabling optimization of the weighting vectors through least-squares regression. This optimization process utilizes simulation data from both a highly accurate stage model and a set of local observers, \FV{i.e.,} $\lbrace O_i\rbrace_{i=1}^n$. Furthermore, leveraging the accurate simulation dataset, \eqref{SectionIII_eq9} is expanded \FV{as:}
\begin{equation*}
\resizebox{\hsize}{!}{$
\underbrace{\begin{pmatrix}
    \hat{q}(2|1) \\ \vdots \\ \hat{q}(N+1|N)
\end{pmatrix}}_{E} \approx \underbrace{\begin{pmatrix}
    \left(\hat{q}_1(2|1) \ \hdots \ \hat{q}_n(2|1)\right)\otimes \chi(p(1)) \\
    \ \ \ \ \ \ \vdots \\
     \left(\hat{q}_1(N+1|N) \ \hdots \ \hat{q}_n(N+1|N)\right)\otimes \chi(p(N))
\end{pmatrix}}_{U}
\underbrace{\begin{pmatrix}
    \Theta_1 \\ \vdots \\ \Theta_n
\end{pmatrix}}_{F},$}
\end{equation*}

\noindent where $N$ represents the length of the dataset. To ensure proper alignment between local observers and their respective positions, constraint conditions are incorporated into the optimization. These constraints ensure that local observers contribute fully (100\%) to approximating the modal dynamics at their designated positions, while their contributions are set to zero elsewhere. This alignment is achieved by:
\begin{equation*}
   \underbrace{\left(\begin{pmatrix}
        \chi(p(1)) \\ \vdots \\  \chi(p(n))
    \end{pmatrix}\otimes I_{n\times n} \right)}_{X} \underbrace{\begin{pmatrix}
    \Theta_1 \\ \vdots \\ \Theta_n
\end{pmatrix}}_{F} = \underbrace{\begin{pmatrix}
    I_{n\times n} (1\ 0\ \hdots \ 0)^\top \\
    \ \ \ \  \vdots \\
    I_{n\times n} (0\ \hdots \  0 \ 1)^\top 
\end{pmatrix}}_{J},
\end{equation*}

\noindent where $X$ encompasses the local position information of the local observers $\lbrace O_i \rbrace_{i=1}^n$ and $J$ enforces the constraint conditions. In this context, the weighting coefficient vectors $\lbrace \Theta_i \rbrace_{i=1}^n$ are obtained by minimizing $\| UF-E \|_2$ subject to $XF=J$. Moreover, through the obtained weighting filter coefficients, the observer is implemented according to \eqref{SectionIII_eq4}. The proposed implementation approach is illustrated in Figure \ref{fig:SectionIII_Fig1} and corresponds to Contribution (C1) of the paper.
\begin{figure}[t]
    \centering
    \includegraphics[trim={2cm 0cm .7cm 0cm},clip,width=\linewidth]{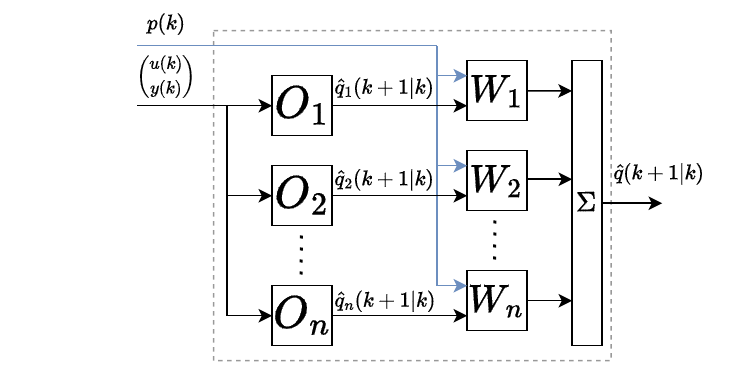}
    \caption{Block interconnection of proposed position dependent observer implementation \FV{for the estimation of a single position dependent flexible mode, where the modal states are approximated through a weighted sum of the outputs of the set of local observers $\lbrace O_i \rbrace_{i=1}^n$}.}
    \label{fig:SectionIII_Fig1}
\end{figure}

For the development of the flexible mode controller $K_{\mathrm{FM}}$, we consider the augmented dynamics denoted by $\mathcal{P}$, as depicted in Figure \ref{fig:Figure1_ASML}, following the control design approach proposed in \cite{Verkerk-phd}.  The primary objective is to construct a modal state feedback $K_{\mathrm{FM}}$ based on the continuous-time (CT) modal dynamics. This approach is advantageous because in the CT scenario, the modal parameters emerge uncoupled in the dynamical equations, facilitating intuitive tuning of the state feedback. Within this framework, the equations of motion expressed by \eqref{EqMOtionSectionII} are expanded by incorporating the additional input from the flexible mode controller:
\begin{equation}
     \Ddot{q}(t) + 2Z\Omega \dot{q}(t) + \Omega^2 q(t) = \tilde{V}^\top \Phi_a\left(u_{\mathrm{RB}}(t)+u_{\mathrm{FM}}(t)\right),
     \label{SectionIII_InfluencedModalDynamics}
\end{equation}

\noindent where the flexible mode control input is defined as:
\begin{equation*}
u_{\mathrm{FM}}(t) = K_{\mathrm{FM}}\begin{pmatrix}
         \hat{q}(t) \\ \dot{\hat{q}}(t)
     \end{pmatrix},  
     \label{SectionIII_FlexModeControlInput}
\end{equation*}

\noindent which can be redefined using the observation error:
\begin{equation*}
    \begin{pmatrix}
         e_o(t) \\ \dot{e}_o(t) 
     \end{pmatrix} = \begin{pmatrix}
         {q}(t) \\ \dot{q}(t)
     \end{pmatrix} - \begin{pmatrix}
         \hat{q}(t) \\ \dot{\hat{q}}(t)
     \end{pmatrix}
\end{equation*}

\noindent as:
\begin{equation}
u_{\mathrm{FM}}(t) = \underbrace{\begin{pmatrix}
    K_{\mathrm{s}} & K_\mathrm{d}
\end{pmatrix}}_{K_{\mathrm{FM}}} \left(\begin{pmatrix}
         {q}(t) \\ \dot{q}(t)
     \end{pmatrix} -  \begin{pmatrix}
         e_o(t) \\ \dot{e}_o(t) 
     \end{pmatrix}\right),  
     \label{SectionIII_EqFM}
\end{equation}

\noindent where the state feedback controller $K_{\mathrm{FM}}$
  is divided into an active stiffness controller $K_\mathrm{s}$
  and an active damping controller $K_\mathrm{d}$. Moreover, substitution of \eqref{SectionIII_EqFM} in \eqref{SectionIII_InfluencedModalDynamics} yields:
\begin{equation}
\begin{split}
        \Ddot{q}(t)+\overbrace{\begin{pmatrix}2Z\Omega-\tilde{V}^\top \Phi_a K_d  \end{pmatrix}}^{2Z_\ast \Omega_\ast}\dot{q}(t) + \overbrace{\begin{pmatrix}\Omega^2 - \tilde{V}^\top \Phi_a K_s \end{pmatrix}}^{\Omega_\ast^2}q(t)  \\
        =\tilde{V}^\top \Phi_a \begin{pmatrix}u_{\mathrm{RB}}(t)-K_se_o(t) - K_d \dot{e}_o(t) \end{pmatrix}
        \end{split} ,
            \label{ModeControlMatrices}
\end{equation}

\noindent where $Z_\ast$ is a diagonal matrix containing the desired modal damping parameters and $\Omega_\ast$ is a diagonal matrix containing the desired eigenfrequencies. These matrices are actively influenced by the flexible mode controller $K_{\mathrm{FM}}$, allowing for intuitive tuning of the modal dynamics, respectively.

From \eqref{ModeControlMatrices} several conclusions are drawn. First, it is observed that integration of flexible mode control loop allows for active manipulation of modal parameters of the moving-body by properly shaping $K_{\mathrm{FM}}$. Secondly, it is observed that the observation error of a specific flexible mode results in additional excitation of the mode and thus further deteriorates position tracking performance. Furthermore, the flexible mode controller $K_{\mathrm{FM}}$ is constructed in a model-based manner as:
\begin{equation}
    \begin{split}
        K_s &= -\begin{pmatrix}\tilde{V}^\top \Phi_a \end{pmatrix}^\dagger \begin{pmatrix}\Omega_\ast^2 - \Omega^2 \end{pmatrix} \\ 
        K_d &= -\begin{pmatrix}\tilde{V}^\top \Phi_a \end{pmatrix}^\dagger\begin{pmatrix}2Z_\ast \Omega_\ast - 2Z\Omega \end{pmatrix}
    \end{split}
    \label{ActiveDampingEquation}
\end{equation}

Moreover, the total design procedure for the flexible mode control loop corresponds to Contribution (C2) of the paper.
\end{subsection}
\end{section}

\begin{section}{Experimental Validation}
\label{ExperimentalVAlidationSection}
\begin{subsection}{System Description}

A state-of-the-art EUV wafer stage system, as shown in Figure \ref{fig:SectionIV_PrototypeFigure}, comprises a dual-stroke mechanism incorporating a balance mass, as detailed in \cite{EUVLITHO}. The long-stroke function of the stage employs a magnetically levitated moving-coil motor controlled by the Lorentz principle, facilitated by the magnet plate. This control mechanism enables precise positioning within the micro-meter range. To achieve accurate wafer positioning, the short stroke (WS-SS) motor is utilized, operating at sub-nanometer levels within the constrained stroke of the long-stroke motor. Interferometry is employed for relative position measurements, establishing position-dependent relationships between measured positions and the control point on the moving body. Control of the WS-SS motor occurs in six degrees of freedom (DoF) through six voice coils, \FV{i.e.,} $x,y,z,R_x,R_y,R_z$. A position-dependent rigid-body decoupling approach is employed for the design of the rigid-body feedback controller \FV{$K_{\mathrm{RB}}$}, facilitating sequential-loop-closing-based motion control design.

    \begin{figure}[t]
        \centering
        \includegraphics[width=\linewidth]{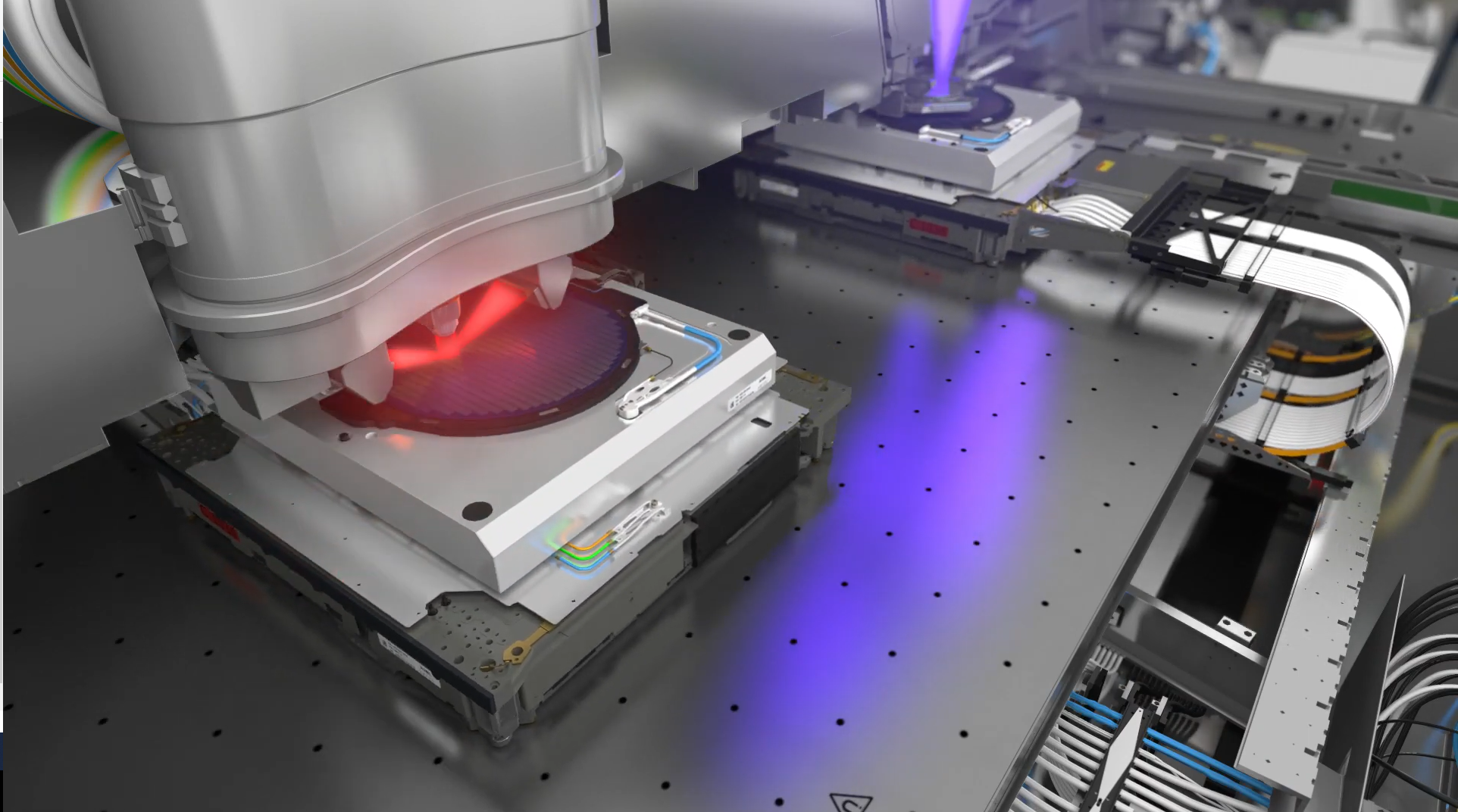}
        \caption{State-of-the-art EUV wafer stage system.}
        \label{fig:SectionIV_PrototypeFigure}
    \end{figure}
\end{subsection}

\begin{figure}[b]
    \centering
    \includegraphics[trim={2.8cm 1.4cm 2.8cm 1.12cm},clip,width=\linewidth]{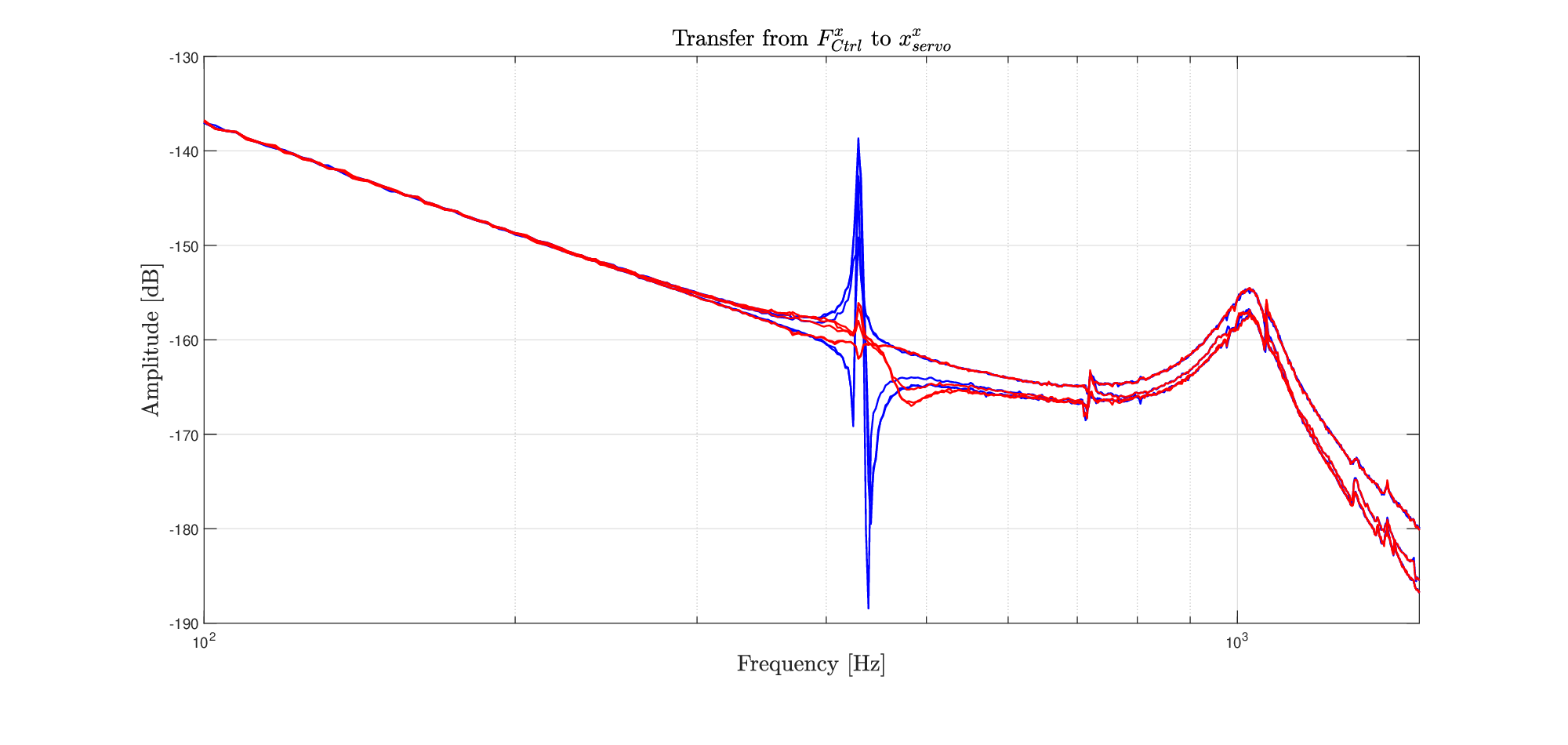}
    \caption{Local FRF measurements of the \emph{equivalent mechanics}  $\mathcal{P}$ of the transfer in $x$-direction ($u_{\mathrm{RB}}^x \rightarrow y_{\mathrm{RB}}^x$), where the blue graph corresponds to local FRF measurements of the conventional rigid body control loop. The red graph denotes to  FRF measurements for which the proposed active damping loop is closed.}
    \label{fig:FRFmeasuremetns}
\end{figure}

\begin{subsection}{Experimental Results}
    To illustrate the functionality of the proposed flexible mode control loop, the control interconnection illustrated by Figure \ref{fig:Figure1_ASML} is implemented on the experimental prototype. It is important to note that as there are only six actuators present on the WS-SS, only six mechanical DoF can be controlled without introducing channel interaction. To avoid interaction between the rigid-body control loop and the flexible mode control loop, a band-pass filter $K_{\mathrm{BP}}$ is integrated in the flexible mode control loop, \FV{i.e.,} $\tilde{K}_{\mathrm{FM}}=K_{\mathrm{FM}}K_{\mathrm{BP}}$,with: 
    \begin{equation*}
    K_{\mathrm{BP}} = \text{diag}\left(\left(\frac{\frac{\omega_i}{Q}s}{s^2+\frac{\omega_i}{Q}s+\omega_i^2}\right)^2 \right) ,
\end{equation*}

\noindent where $s$ denotes the complex frequency, $\omega_i$ is the eigenfrequency of the $i$-th flexibly mode and $Q$ corresponds to a tuning parameter that regulates the filter band. Integration of this filter in the flexible mode control loop avoids interaction between the rigid-body modes and the flexible mode control loop.

To evaluate the effectiveness of our approach, we conducted two types of experiments. Firstly, we controlled the first flexible mode of the system using an active damping controller $K_\mathrm{d}$, constructed using Equation \eqref{ActiveDampingEquation}. We investigated the impact of the flexible mode control loop on the augmented dynamics $\mathcal{P}(p(t))$, as perceived by the rigid-body feedback controller. Results \FV{for 9 local positions of the moving-body }are presented in Figure \ref{fig:FRFmeasuremetns}, showing local Frequency Response Functions (FRFs) of $\tilde{P}(p(t))$ in blue and local FRFs of $\mathcal{P}(p(t))$ in red for the transfer in the rigid-body x-direction. Our findings indicate that implementing the proposed active damping loop facilitates desirable manipulation of the plant from the perspective of the rigid-body feedback controller, thus easing motion control design constraints. Moreover, closure of the flexible mode control loop leads to approximately an 18 dB suppression of the resonance mode, as observed.

\begin{figure}[t]
    \centering
    \includegraphics[trim={0.3cm 0cm 0.3cm 0cm},clip,width=\linewidth]{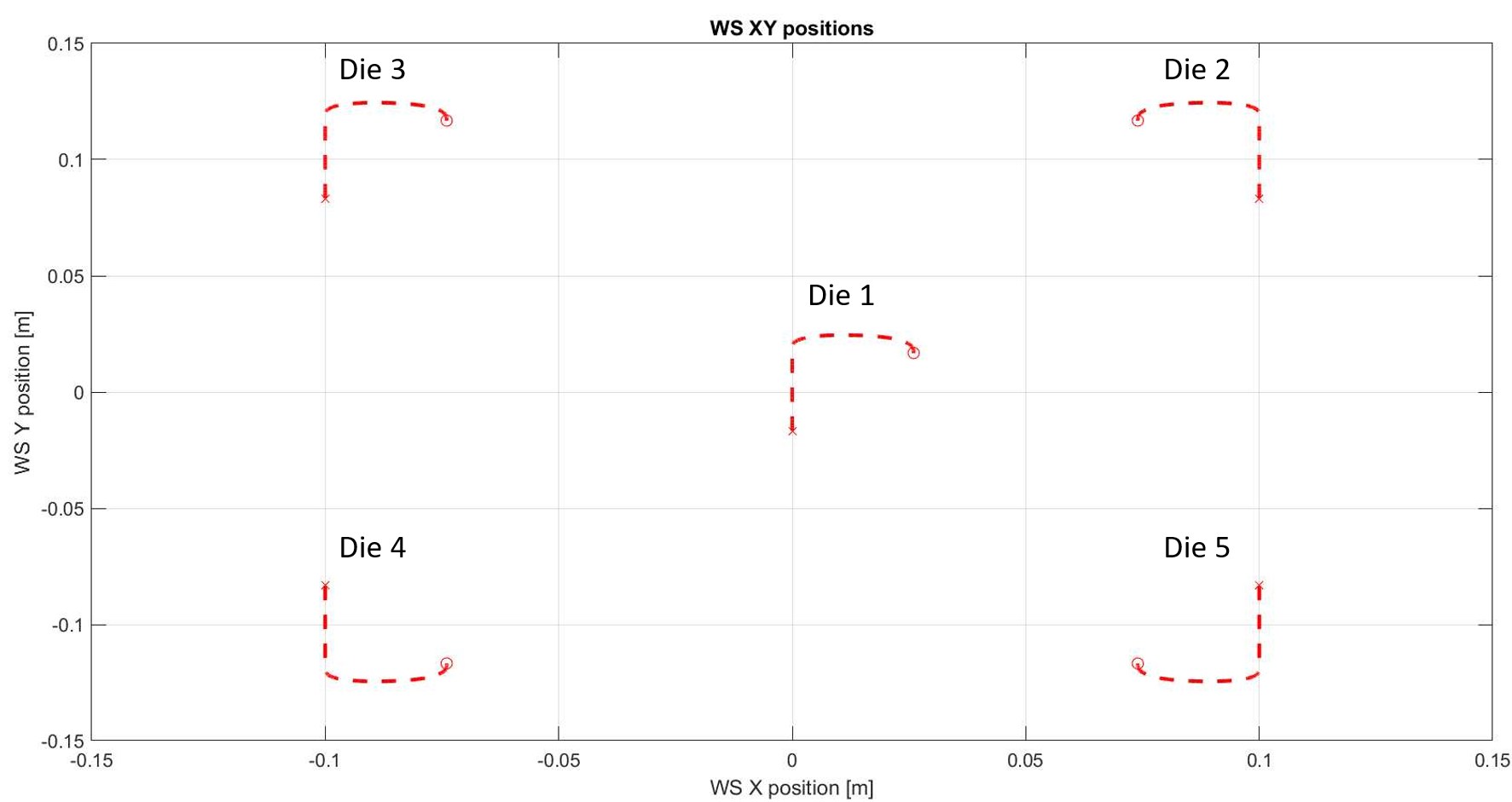}
    \caption{Five motion profiles that are considered to evaluate improvement of position tracking performance using the proposed control approach.}
    \label{fig:referenceprofiles}
\end{figure}

Secondly, time-domain measurements were conducted to assess the impact of the active damping loop on the position tracking performance of the system. This involved a comparison between the currently employed state-of-the-art rigid-body control configuration and the proposed extension, which actively dampens the first resonance mode. For both cases, the feedback controller $K_{\mathrm{RB}}$ and the feedforward controller $K_{\mathrm{FF}}$ have been chosen identical. \FV{For the design of the flexible mode controller, 9 local observers were designed using the approach presented in Section \ref{Section:ControlDesignApproach}, where for experimental validation $m_x$, $m_y$ were considered to be 1.}

For experimental validation, five motion profiles were considered, as depicted in Figure \ref{fig:referenceprofiles}. These motion profiles were designed using the 4th-order reference trajectory generator developed by \cite{208Lambrechts}, with parameters set as follows: $a_x^{\mathrm{max}}=35 \frac{m}{s^2}$, $a_y^{\mathrm{max}}=15 \frac{m}{s^2}$, $v_x^{\mathrm{max}}=0.8 \frac{m}{s}$, and $v_y^{\mathrm{max}}=0.38 \frac{m}{s}$\FV{, where $m$ corresponds to meter and $s$ denotes seconds.}

In the semiconductor industry, assessing position tracking accuracy occurs during the exposure of the silicon wafer, particularly during the constant velocity interval. The system's performance is evaluated using two metrics: the Moving Average (MA) and the Moving Standard Deviation (MSD), see \cite{176Butler}. These metrics are computed from the raw position tracking error as follows:
\begin{equation*}
    \begin{split}
        \mathrm{MA(t)} &= \frac{1}{T}\int_{t-\frac{T}{2}}^{t+\frac{T}{2}} e(\tau)d\tau\\
        \mathrm{MSD(t)} &= \sqrt{\frac{1}{T}\int_{t-\frac{T}{2}}^{t+\frac{T}{2}} (e(\tau)-MA(t))^2d\tau}
    \end{split},
\end{equation*}

\noindent where $T$ corresponds to the exposure time interval.

\begin{figure}[t]
    \centering
    \includegraphics[width=0.7\linewidth]{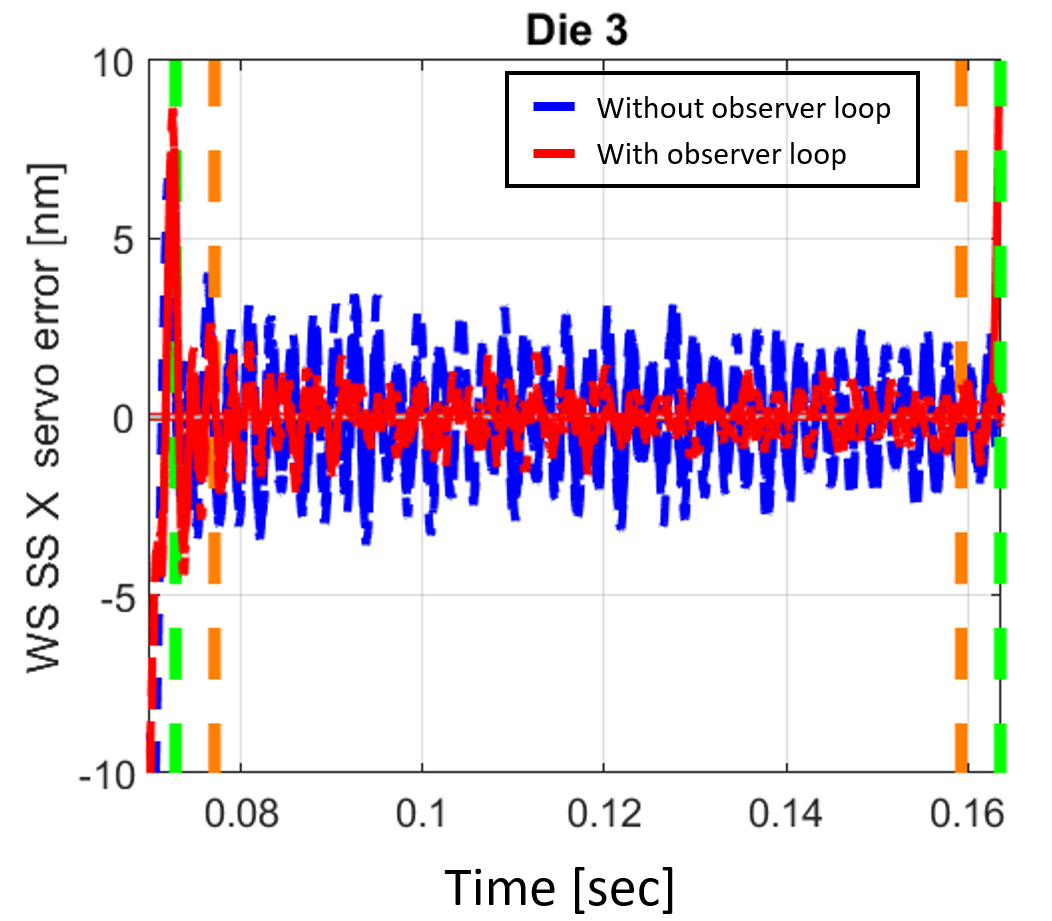}
    \caption{Raw position tracking error of Die 3, where the blue graph corresponds to the conventional rigid body control structure and the red graph corresponds to the proposed control approach. The green window indicates the constant velocity interval and the orange window indicates the exposure interval.}
    \label{fig:servoposerror}
\end{figure}

\begin{figure}[b]
    \centering
    \includegraphics[width=0.7\linewidth]{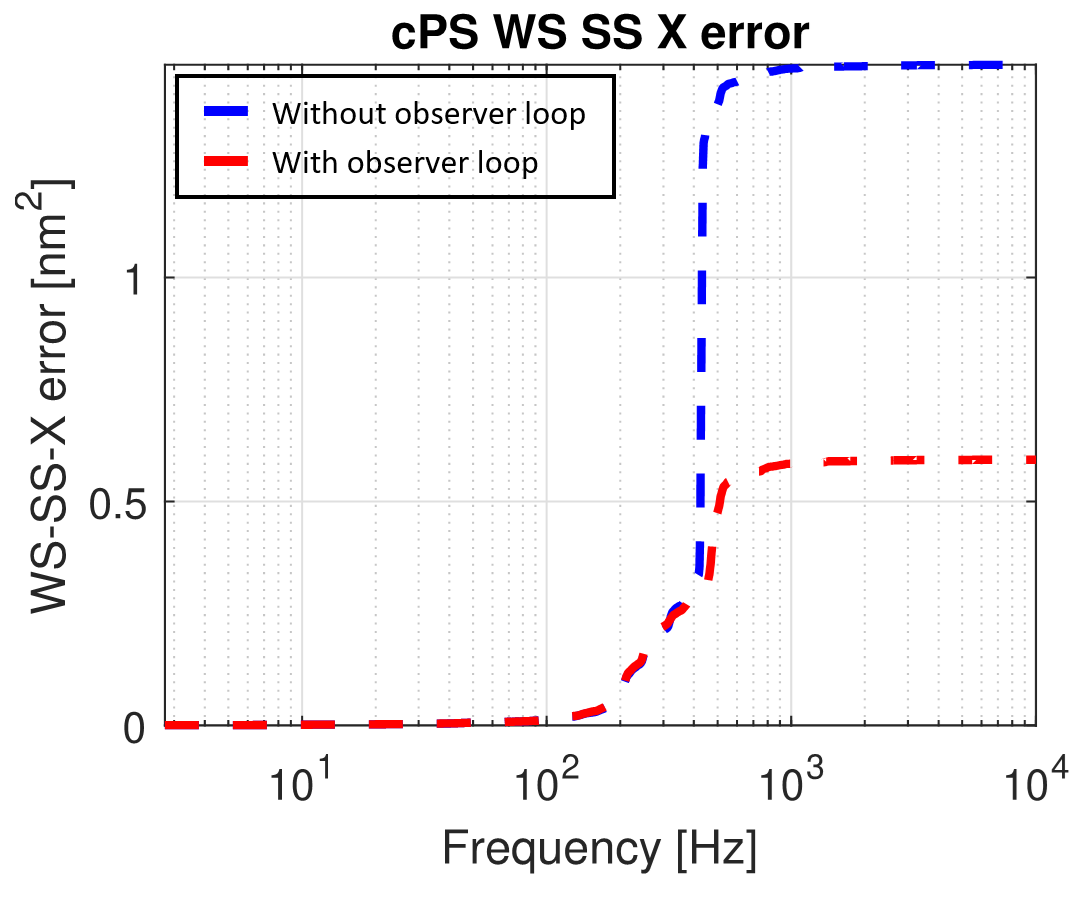}
    \caption{Cumulative power spectral density of the raw position tracking error in $x$-direction ($u_{\mathrm{RB}^x} \rightarrow y_{\mathrm{RB}}^x$), where the blue graph corresponds to the cPS of the conventional rigid body control structure and the red graph denotes the cPS of the proposed control approach.}
    \label{fig:cPS}
\end{figure}

Moreover, Figure \ref{fig:servoposerror} depicts the raw position tracking error in the $x$-direction ($u_{\mathrm{RB}}^x \rightarrow y_{\mathrm{RB}}^x$) of Die 3. The blue graph represents the position tracking error of the conventional rigid body feedback control loop, while the red graph illustrates the position tracking error of the proposed control approach. Additionally, the green window signifies the constant velocity time interval, and the orange window indicates the exposure time interval. By observing Figure \ref{fig:servoposerror}, it is evident that the proposed control approach enhances position tracking performance compared to the conventional rigid body control structure by actively damping the first resonance mode. This observation is supported by the cumulative power spectral density (cPS) of the raw position tracking error during the exposure time interval, as shown in Figure \ref{fig:cPS}.

The analysis from Figure \ref{fig:cPS} indicates that introducing an active damping loop to the rigid body control structure significantly diminishes the contribution of the resonance mode towards the position tracking error. To quantify this improvement, the MSD performance of both control structures is evaluated, as summarized in Table \ref{MSDPerformance}. Table \ref{MSDPerformance} demonstrates that the proposed control structure significantly enhances MSD performance across all six physical axes. This underscores the effectiveness of the proposed control design approach in achieving improved position tracking performance for high-precision motion systems.

\begin{table}[t]
\begin{center}
\begin{tabular}{|l|l|l|}
\hline
                     & Conventional control  & Proposed control\\ 
     & structure &     structure           
                     \\
                     \hline
 $x$ {[}nm{]}     & 2.32                            & 1.51                         \\ \hline
$y$ {[}nm{]}     & 0.86                            & 0.71                         \\ \hline
$z$ {[}nm{]}     & 6.77                            & 2.25                         \\ \hline
$R_x$ {[}nrad{]} & 14.89                           & 6.64                         \\ \hline
$R_y$ {[}nrad{]} & 52.88                           & 16.83                        \\ \hline
$R_z$ {[}nrad{]} & 65.39                           & 33.10                        \\ \hline
\end{tabular}
\caption{MSD performance values of the original control structure and the MSD performance values of the proposed control structure.}
\label{MSDPerformance}
\end{center}
\end{table}

\end{subsection}
\end{section}

\begin{section}{Conclusions}
\label{Section:Conclusion}
    This paper presents a novel output-based modal observer design approach to actively manage flexible dynamics. To facilitate real-time implementation, we introduce an implementation strategy leveraging the position-dependent characteristics of the plant. This strategy incorporates suitable position-dependent weighting functions, enabling a computationally efficient implementation of the control approach. Experimental validation on a cutting-edge EUV wafer stage, which manifests position-dependent effects in the position measurements of the moving-body, demonstrates the efficacy of our proposed control approach in enhancing position tracking performance for high-precision motion systems. Moreover, the introduction of the flexible mode control loop brings several advantages. Firstly, it diminishes the contribution of flexible dynamics to position tracking errors. Secondly, by actively damping resonance dynamics, it allows for increased bandwidth of rigid body feedback controller, thereby enhancing disturbance rejection capabilities. 
\end{section}
%In this section, the extension of the conventional rigid body control structure towards active control of flexible dynamics is presented. For the design of the position-dependent modal control loop, a two step approach is considered. First, the position-dependent observer and the static state feedback are designed independently. Secondly, closed-loop stability is investigated by showing existence of a Lyapunov function, see \cite{scherer2000linear}.

%This Section is organized as follows. Subsection \ref{subsection:observer} presents the design of the position dependent modal observer $O$, which reconstructs the internal dynamics of the moving-body. Moreover, Subsection \ref{subsection:feedback} presents the design of the modal state feedback $K_{FM}$, which allows for active manipulation of the modal parameters of the moving-body respectfully. At last, closed-loop stability is investigated in Subsection \ref{subsection:Closed-loop-stability}.

%%%%%%%%%%%%%%%%%%%%%%%%%%%%%%%%%%%%%%%%%%%%%%%%%%%%%%%%%%%%%%%%%%%%

%%%%%%%%%%%%%%%%%%%%%%%%%%%%%%%%%%%%%%%%%%%%%%%%%%%%%%%%%%%%%
%===========================================================================================

%=================================================================================
%\input{Appendix}

%===========================================================================================

\bibliographystyle{ieeetr}       
\bibliography{MyBib}

%===========================================================================================

\end{document}